\documentclass[12pt]{article}
\usepackage{amssymb}

\newcommand{\tu}{\tilde{U}}
\newcommand{\tv}{\tilde{V}}
\newcommand{\be}{\begin{equation}}
\newcommand{\ee}{\end{equation}}
\newcommand{\bn}{\begin{eqnarray}}
\newcommand{\en}{\end{eqnarray}}
\newcommand{\p}{\partial}

\newcommand{\as}{A\!\!\!/}

\newcommand{\sr}{\sum_{r=1}^N}
\newcommand{\ssn}{\sum_{s=1}^N}
\newcommand{\nn}{\nonumber}

\newcommand{\no}{\noindent}

\def\bea{\begin{eqnarray}}
\def\eea{\end{eqnarray}}
\begin{document}
\title{\textbf{Multiflavor Soldering}}
\author{D. Dalmazi and A. de Souza Dutra  \\
\textit{UNESP - Campus de Guaratinguet\'a - DFQ } \\
\textit{Av. Dr. Ariberto Pereira da Cunha, 333} \\
\textit{CEP 12516-410 - Guaratinguet\'a - SP - Brazil.} \\
\textsf{E-mail: dalmazi@feg.unesp.br, dutra@feg.unesp.br }} \maketitle

\begin{abstract}

In two dimensions the simple addition of two chiral bosons of
opposite chiralities does not lead to a full massless scalar
field. Similarly, in three dimensions the addition of two
Maxwell-Chern-Simons fields of opposite helicities $\pm 1$ will
not produce a parity invariant Maxwell-Proca theory. An
interference term between the opposite chiralities (helicities)
states is required in order to obtain the expected result. The so
called soldering procedure provides the missing interference
Lagrangian in both 2D and 3D cases. In two dimensions such
interference term allows to fuse two chiral fermionic determinants
into a nonchiral one.  In a recent work we have generalized this
procedure by allowing the appearance of an extra parameter which
takes two possible values and leads to two different soldered
Lagrangians. Here we apply this generalized soldering  in a
bosonic theory which has appeared in a partial bosonization of the
$3D$ gauged Thirring with $N$ flavors. The multiplicity of flavors
allow new types of solderings and help us to understand the
connection between different perturbative approaches to
bosonization in $3D$. In particular, we obtain an interference
term which takes us from a multiflavor Maxwell-Chern-Simons theory
to a pair of self-dual and anti-self-dual theories when we combine
together both fermionic determinants of $+1/2$ and $-1/2$ helicity
fermions. An important role is played by a set of pure
non-interacting Chern-Simons fields which amount to a
normalization factor in the fermionic determinants and act like
spectators in the original theory but play an active role in the
soldering procedure. Our results suggest that the generalized
soldering could be used to provide dual theories in both $2D$ and
$3D$ cases.

\end{abstract}

\section{Introduction}

A massless scalar field in $D=1+1$ is one of the simplest examples
of a field theory and it is known for a long time to consist of
two chiral fields (left- and right-mover). However, it has taken
some time until one could formulate a consistent field theory for
each chirality \cite{siegel,fj}. An interesting point is to
understand how to recover the massless scalar field from two
chiral bosons of opposite chiralities. This question has been
addressed in \cite{ms,adw,abw,aw}, see also \cite{topics}. It
turns out that the simple addition of the chiral fields does not
produce the desired result, an interference term is necessary, an
explicit expression can be found in \cite{aw}. From the fermionic
point of view one can say that the soldering formalism, with the
help of a left/right symmetry requirement, furnishes the missing
counterterm necessary to combine together two chiral determinants
into a nonchiral one \cite{abw}. This picture extends to the
nonabelian case, see appendix of \cite{abw}. Similarly, in $D=2+1$
the addition of two Maxwell-Chern-Simons (MCS) theories
representing free massive particles of opposite helicities $\pm 1$
does not lead to a parity invariant Maxwell-Proca theory with the
same spectrum, once again an interference term is required. By
using in essence the same procedure of \cite{ms,adw,abw,aw} the
authors of \cite{bw} have been able to provide the missing
interference Lagrangian. The role of the chiral determinants is
now played by the determinants of two components fermions of
helicity $+1/2$ and $-1/2$. The states of helicity $\pm 2$ which
appear in the linearized Einstein-Hilbert-Chern-Simons gravity in
$D=2+1$ have also been joined together in an analogous fashion
\cite{iw}. Such procedure has been called soldering in the
literature and consists of lifting a global shift symmetry into a
local one with the addition of auxiliary fields as we explain in
section 2. In \cite{dda}, we have generalized this procedure by
introducing an extra parameter in the local symmetry which allows,
for instance, the soldering of two MCS theories of different
masses into one MCS-Proca theory with the same spectrum, see also
\cite{bk}. In the present work the use of this generalized
soldering will be essential. This is to the best we know the first
attempt to apply the soldering ideas to multiflavor theories and
leads to new possibilities of parallel and cross-soldering amongst
different flavors. In particular, the use of a spectator field
indicates the existence of a relationship between generalized
soldering and dual theories which might be explored in different
dimensions. The soldering procedure can also be used to unravel
the spectrum of the model avoiding field redefinitions. In the
next section we start from bosonic Lagrangians coming from a
partial bosonization, see \cite{ddh}, of the gauged Thirring model
with $N$ two component fermions of well defined helicities and
explain the logic of the soldering mechanism. In the third section
we make field redefinitions which help us to understand the
soldered Lagrangians of the second section. In the fourth section,
by recalling the twofold generalized soldering of Chiral Schwinger
models of opposite chiralities we are led to conjecture a relation
between generalized soldering and dual theories. At the end
section we draw some conclusions. In the appendix we work out
explicitly the case of $N=3$ flavors in order to convince the
reader about the general rule for the soldered Lagrangians for
arbitrary $N$ flavors.

\section{Twofold Soldering}
In \cite{ddh} we have carried out a partial bosonization of
two-component massive fermions minimally coupled to a gauge field
($QED_{2+1}$) plus a Thirring term:

\be {\cal L}_F^{(+)} = \sum_{r=1}^N \bar{\psi}_{r}\,(i\,\partial \!\!\!/\,-\,m\,-\,
\frac{e}{\sqrt{N}}\,\as)\psi_{r} - \,\frac{g^{2}}{2N}\,(\sum_{r=1}^N \, \overline{\psi }_{r}\gamma ^{\mu }\psi
_{r})^{2}\, -\frac{1}{4}\,F_{\mu \nu }^{2}(A) \label{lmais}\ee

\no Although we have only worked with $+ 1/2$ helicity fermions\footnote{In $2+1$ dimensions
 if we work with two-component spinors the two by two gamma matrices satisfy the algebra $
\left\lbrack \gamma_{\mu},\gamma_{\nu} \right\rbrack = 2 \epsilon_{\mu\nu\alpha} \gamma^{\alpha}$ and
consequently \cite{jn} the operator $S_{\mu}= \gamma_{\mu}/2$ obeys the angular momentum algebra and plays  the
role of the fermion spin. So the Dirac equation $\bar{\psi}\,(i\,\p \!\!\!/\,\mp\,m)\psi =0 $ can be written as
$\left( S_{\mu}P^{\mu} \mp m/2 \right) \psi = 0 $ assuring that massive two-component fermions have helicity $
S\cdot P / m = \pm 1/2$ in $2+1$ dimensions.}  we could have equally started with $-1/2$ helicity states, i.e.,

\be {\cal L}_F^{(-)} = \sum_{r=1}^N \bar{\chi}_{r}\,(i\,\partial \!\!\!/\,+\,m\,-\,
\frac{e}{\sqrt{N}}\,\as)\chi_{r} - \,\frac{g^{2}}{2N} \,(\sum_{r=1}^N \,\overline{\chi }_{r}\gamma
^{\mu}\chi_{r})^{2}\, - \frac{1}{4}\,F_{\mu \nu }^{2}(A) \label{lmenos}\ee

In \cite{ddh} an auxiliary vector field $B_{\mu}$ has been
introduced  in order to lower the quartic Thirring vertex to a
cubic vertex. Then, after introducing sources for the gauge field
and each fermion current $j_{\mu (r)}^{(+)}= \bar{\psi}_{r}
\gamma_{\mu} \psi_{r}$ we have integrated over the fermions at
leading order in $1/N$ and obtained a quadratic action in the
sources and fields $A_{\mu}$ and $ B_{\mu}$.  After Gaussian
integrating the auxiliary field $B_{\mu}$ we have computed two
point correlation functions $\left\langle j_{\mu (r)}^{(+)}(k)
j_{\nu (s)}^{(+)}(-k)\right\rangle $ , $\left\langle j_{\mu
(r)}^{(+)}(k) A_{\nu}(-k)\right\rangle $ and $\left\langle
A_{\mu}(k) A_{\nu}(-k)\right\rangle $. We have suggested a
bosonization map for the fermionic currents $j_{\mu (r)}^{(+)}$
and for the Lagrangian ${\cal L}_F^{(+)}$ compatible with those
specific\footnote{Two point functions of other operators like the
mass term $\left\langle \bar{\psi}\psi (k)
\bar{\psi}\psi(-k)\right\rangle$ have not been considered in the
above maps.} two point functions. The map is independent of the
Thirring and QED couplings but it is in general nonlocal. However,
in the large fermion mass limit ($m\to\infty$), assumed here
henceforth, it becomes local. If we include now the $-1/2$
helicity fermions we can write down the map of \cite{ddh} in terms
of $2N$ bosonization\footnote{In this work there is no sum over
repeated flavor indices unless otherwise stated.} fields
$U_r^{\gamma},V_r^{\gamma}$ :

\be \bar\psi_r \gamma_{\mu} \psi_r = \epsilon_{\mu\nu\gamma} \p^{\nu} U^{\gamma }_r \quad ; \quad \bar\chi_r
\gamma_{\mu} \chi_r = \epsilon_{\mu\nu\gamma} \p^{\nu} V^{\gamma }_r \label{jmap} \ee

\be \bar{\psi}_{r}\left(\,\partial \!\!\!/\,-\,m \right)\psi_{r} = 2 \pi
\epsilon_{\mu\nu\gamma}U^{\mu}_r\p^{\nu} U^{\gamma}_r \label{lbmais} \ee

\be \bar{\chi}_{r}\left(\,\partial \!\!\!/\,+\,m \right)\chi_{r} = - 2 \pi
\epsilon_{\mu\nu\gamma}V^{\mu}_r\p^{\nu} V^{\gamma}_r \label{lbmenos} \ee

\no The bosonized versions of the actions $\int d^3 x {\cal L}_F^{\pm}$, except for the Maxwell term,  are given
respectively by :

\be W_+ = \int d^3 x \left\lbrack 2\pi\, \sum_{r=1}^N \epsilon_{\mu\nu\gamma}U_r^{\mu}\p^{\nu}U_r^{\gamma} -
\frac{ i \, e}{\sqrt{N}} A^{\mu} \sum_{r=1}^N \epsilon_{\mu\nu\gamma}\p^{\nu}U_r^{\gamma} - \frac{g^2}{4 N}
\left( \sum_{r=1}^N F_{\mu\nu} ( U_r ) \right)^2 \right\rbrack \label{sbmais}\ee

\be W_- = \int d^3 x \left\lbrack - 2\pi\, \sum_{r=1}^N \epsilon_{\mu\nu\gamma}V_r^{\mu}\p^{\nu}V_r^{\gamma} -
\frac{ i \, e}{\sqrt{N}} A^{\mu} \sum_{r=1}^N \epsilon_{\mu\nu\gamma}\p^{\nu}V_r^{\gamma} - \frac{g^2}{4 N}
\left( \sum_{r=1}^N F_{\mu\nu} ( V_r ) \right)^2 \right\rbrack \label{sbmenos}\ee

\no We stress that the bosonic maps described by the formulas (\ref{jmap}-\ref{sbmenos}) hold only at
$m\to\infty$ and at the quadratic approximation for the fermionic determinant which amounts to consider the
vacuum polarization diagram. The actions $W_{\pm}$ are invariant under rigid translations $\delta U_r^{\mu} =
\eta^{\mu}_r \, , \, \delta V_r^{\mu} = \tilde{\eta}^{\mu}_r $. The basic idea of the soldering formalism
\cite{adw,abw} is to combine $W_+$ and $W_-$ into one theory which depends only on a combination of the fields
$U_r^{\mu}$ and $V_r^{\mu}$ . This is realized by promoting the rigid translations to a local symmetry. Namely,
let us suppose the local transformations:

\be \delta U_r^{\mu} = \eta^{\mu}_r \quad ; \quad \delta V_r^{\mu} = \alpha_r \eta^{\mu}_r \label{deltauv}\ee

\no where $\alpha_r$ are so far arbitrary constants. Under (\ref{deltauv}) we have:

\be \delta ( W_+ + W_-) = \sum_{r=1}^N \int d^3x J_{\mu\nu (r)} \p^{\mu}\eta^{\nu}_r \label{deltaw} \ee

\no where

\be J_{\mu\nu (r)} = - \frac{g^2}N \sum_{s=1}^N \left( F_{\mu\nu}(U_s) + \alpha_r F_{\mu\nu}(V_s )\right) + 4\pi
\epsilon_{\mu\nu\gamma}\left(U_r^{\gamma} - \alpha_r V^{\gamma}_r\right) - \frac{ e}{\sqrt{N}}
\epsilon_{\mu\nu\gamma}(1 + \alpha_r)A^{\gamma} \label{jmn} \ee

\no Now we introduce $N$  antisymmetric auxiliary fields ($B^{\mu\nu}_r = - B^{\nu\mu}_r$) such that:

\be \delta B^{\mu\nu}_r = - \frac{\left(\p^{\mu}\eta^{\nu}_r - \p^{\nu}\eta^{\mu}_r\right)}2 \label{deltabmn}
\ee

\no If we assume that $\alpha_r^2 = 1 \, , \, r=1,2,\cdots N$, the variation $\delta J_{\mu\nu (r)} $ will
depend only on derivatives of the local parameter $\eta_r^{\mu}$. Consequently we have

\bea \delta  \left( W_+ + W_-  + \int d^3 x \sr B^{\mu\nu}_r J_{\mu\nu (r)}\right) &=& \int d^3 x \sr
B^{\mu\nu}_r \delta
J_{\mu\nu (r)} \nn\\
&=& \int d^3 x \sr B^{\mu\nu}_r \left\lbrack - \frac{g^2}N \ssn ( 1 + \alpha_s\alpha_r
)F_{\mu\nu}(\eta_s)\right\rbrack \nn\\
&=& \frac{g^2}N \int d^3 x \ssn \sr 2\, B^{\mu\nu}_r \, \delta B_{\mu\nu\ (s)} ( 1  + \alpha_s\alpha_r ) \nn \\
\eea

\no Therefore, the soldered  action invariant under (\ref{deltauv}) and (\ref{deltabmn}) is given by:

\be W^{(S)} = \int d^3 x {\cal L}^{(s)} = W_+ + W_- + \int d^3 x \left\lbrack  \sr B^{\mu\nu}_r J_{\mu\nu (r)} -
\frac{g^2}{N} \sr\ssn ( 1 + \alpha_s\alpha_r)B^{\mu\nu}_r B_{\mu\nu (s)}\right\rbrack  \label{ls} \ee

\no Since we have two choices for each parameter $\alpha_r =\pm 1$ we have in principle several possible
soldered Lagrangians. In what follows we eliminate the auxiliary fields $B^{\mu\nu}_r$ through their equations
of motion and show that we end up with only two possibilities. The elimination of auxiliary fields is more
subtle now than in the one-flavor case treated in \cite{bw,dda}. We illustrate the general case of $N$ flavors
by considering first $N=2$. In this case the piece of ${\cal L}^{(S)}_{N=2}$ which depends on the auxiliary
fields is given by ( below $B_r \cdot J_s = B^{\mu\nu}_r J_{\mu\nu (s)} \, , \, B_r \cdot B_s = B^{\mu\nu}_r
B_{\mu\nu (s)} $ ):

\be {\cal L}^{(S)}_{N=2} (B^{\mu\nu}_r) = B_1 \cdot J_1 +  B_2 \cdot J_2 - g^2 \left\lbrack B_1\cdot B_1 +
B_2\cdot B_2 +
  (1 + \alpha_1\alpha_2)\, B_1 \cdot B_2   \right\rbrack \label{l2a} \ee

\no The first case  $\alpha_1=\alpha_2=\alpha $  implies:

\be {\cal L}^{(S)}_{N=2} (B^{\mu\nu}_r) = B_- \cdot J_- +  B_+ \cdot J_+ - g^2 \left( B_+\cdot B_+
\right)\label{l2b} \ee

\no Where $B_{\pm}^{\mu\nu} = B^{\mu\nu}_1 \pm B^{\mu\nu}_2 \, , \, J_{\pm}^{\mu\nu} = \left( J^{\mu\nu}_1 \pm
J^{\mu\nu}_2 \right)/2 $. The equations of motion for  $B_{\mu\nu(-)}$
 lead to the identification $J_{\mu\nu
(1)}=J_{\mu\nu (2)}\equiv J_{\mu\nu (\alpha)}=J_{\mu\nu (+)}$.
 Thus, eliminating $B_{\mu\nu(+)}
$ we end up  with the interference Lagrangian density

\be {\cal L}_{N=2}^{I}(\alpha,\alpha) = \left( J_{\alpha}\cdot J_{\alpha} \right)/(4 g^2)\label{laa} \ee

\no In the second case where $\alpha_1= -\alpha_2 = \alpha $ in (\ref{l2a}) the auxiliary fields decouple. Now
we have, after elimination of such fields, a different interference Lagrangian density

\be {\cal L}_{N=2}^{I}(\alpha,-\alpha) = \left(J_{\alpha}\cdot J_{\alpha} + J_{-\alpha}\cdot J_{-\alpha}
\right)/(4 g^2) \label{lama} \ee

\no Where $J_{\mu\nu (1)}= J_{\mu\nu (\alpha)}$ and $J_{\mu\nu (2)}= J_{\mu\nu (-\alpha)}$ In the appendix we
work out explicitly the case $N=3$. It is not difficult to convince oneself that for $N$ flavors we still have
only two possibilities for interference Lagrangians, namely, either we have all the constants $\alpha_r$ with
the same sign or at least one of them with a different sign. We start with the second case, if we have $M$
constants of the $\alpha$-type ($\alpha_i=\alpha, \, i=1,\cdots, M$) and $N-M$ constants of the opposite sign
($\alpha_k=-\alpha, \, k=M+1,\cdots, N$), then the elimination of the auxiliary fields $B_{\mu\nu (r)}$ will
lead, if $1\le M \le N -1 $,  to the identifications

\be J_{\mu\nu (1)}=J_{\mu\nu (2)}= \cdots = J_{\mu\nu (M)} = J_{\mu\nu (\alpha)} \label{id1} \ee

\be J_{\mu\nu (M+1)}=J_{\mu\nu (M+2)}= \cdots = J_{\mu\nu (N)} = J_{\mu\nu (-\alpha)} \label{id1b} \ee

\no and to the soldered action :

\be W^{(S)}_{(2)} = W_+ + W_-  + \frac{N}{8 g^2} \int d^3 x \left\lbrack J_{\alpha}\cdot J_{\alpha} + J_{-\alpha
}\cdot J_{-\alpha } \right\rbrack \quad , \label{ws1} \ee

\no The subscript (2) labels the second case. It turns out that we are able to write down $W^{(S)}_{(2)}$
entirely in terms of the combinations $C_r^{\mu} \equiv U^{\mu}_r - \alpha_r V^{\mu}_r $ which are invariant
under the local translations (\ref{deltauv}). The identification of two currents, say $J_{\mu\nu (r)} =
J_{\mu\nu (s)} $ for which $\alpha_r = \alpha_s$ imply, see (\ref{jmn}),  the identification of the combination
of fields $ C_r^{\mu}= C^{\mu}_s $. Therefore, (\ref{id1}) and (\ref{id1b}) allow us to write down
$W^{(S)}_{(2)}$ in terms of only two combinations of fields $C^{\mu}_{(\alpha)}\equiv U^{\mu}_r - \alpha_r
V^{\mu}_r
  $ and $C^{\mu}_{(-\alpha)}\equiv U^{\mu}_s -
\alpha_s V{\mu}_s  $ where $r$ and $s$ may be any flavor in the ranges $1 \le r \le M$ and $M+1 \le s \le N$
respectively. After some rearrangements we have a rather simple soldered Lagrangian independent of $M$, namely:

\bea {\cal L}^{(S)}_{(2)} &=& - (2\pi \, N) \,\epsilon_{\mu\nu\gamma} C^{\mu}_+ \p^{\nu}C^{\gamma}_+ + (2\pi
\, N) \epsilon_{\mu\nu\gamma} C^{\mu}_- \p^{\nu}C^{\gamma}_- \nn\\
&+& \frac 1{2 g^2} \left( e\, A^{\mu} - \, 4\pi \, \sqrt{N} C_+^{\mu} \right)^2 + \frac 1{2 g^2} \left( e\,
A^{\mu} - \, 4\pi \, \alpha \, \sqrt{N} C_-^{\mu} \right)^2 \quad , \label{ls2} \eea

\no where $C_{\pm}^{\mu} = (C^{\mu}_{(\alpha)} \pm C^{\mu}_{(-\alpha)})/2 $. Therefore, the soldering procedure
has led us to two  self-dual models \cite{tpn} of opposite helicities $\pm 1$ and mass $ 2 \pi /g^2 $, linearly
coupled to a gauge field.

In the first case where all constants are equal $\alpha_1=\alpha_2= \cdots = \alpha_N = \alpha $  , i.e., $M=N$
all the currents must be identified

\be J_{\mu\nu (1)}=J_{\mu\nu (2)}= \cdots = J_{\mu\nu (N)} = J_{\mu\nu (\alpha)} \label{id2} \ee

\no which leads to a different soldered action :

\be W^{(S)}_{(1)} = W_+ + W_-  + \frac{N}{8 g^2} \int d^3 x J_{\alpha} \cdot J_{\alpha} \label{ws1} \ee

\no Thus, we have only one type of independent combination of fields, i.e., $C_{(\alpha)}^{\gamma}$ which may be
identified with any combination $C^{\gamma}_r$ with $1\le r \le N$. After some manipulations we have:

\be {\cal L}^{(S)}_{(1)} = -\frac{g^2 N}8 F_{\mu\nu}^2(C_{(\alpha )}) + \frac 1{g^2} \left( e\, A^{\mu} -  2\pi
\, \sqrt{N} C_{(\alpha)}^{\mu} \right)^2 \quad . \label{ls1} \ee

\no So we have now a Maxwell-Proca theory  linearly coupled to a gauge field. It is known \cite{bk,jhep2} that
the Maxwell-Proca theory is dual to a couple of Maxwell-Chern-Simons models of opposite helicities $\pm 1$ with
mass $2\pi/g^2 $. Since each Maxwell-Chern-Simons theory is dual \cite{dj} to a self-dual model we conclude that
the two soldering procedures have furnished dual theories with the same spectrum. Moreover, the gauge invariance
$A_{\mu} \to A_{\mu} + \p_{\mu}\phi $ of the original theories $W_{\pm}$ is still present in both soldered
theories (\ref{ls1}) and (\ref{ls2}) if we transform the combinations $C_{(\alpha)}^{\mu} $ and
$C_{(-\alpha)}^{\mu}$ accordingly. Last, we notice that in (\ref{ls2}) we had to split the $N$ flavors in two
sets of opposite signs and consequently the soldering mechanism has broken the permutation symmetry of the
flavors while in (\ref{ls1}), due to the identification $C_{(r)}^{\mu} = C_{(s)}^{\mu} (1\le r,s \le m)$ we
could replace $C_{(\alpha)}^{\mu}$ by $\sum_{r=1}^N C_r^{\mu}/N $ and keep explicitly the symmetry under
permutation of the flavor indices.

\subsection{Field redefinition}

We start this section by making some field redefinitions in the starting theories $W_{\pm}$ in order to figure
out why we have ended up with only two soldering possibilities containing just one couple of massive states of
opposite helicities $\pm 1$. First of all, for simplicity, we neglect the gauge field. In this case we can
rewrite (\ref{sbmais}) and (\ref{sbmenos}) as follows:

\be W_+ = \sum_{r=1}^N \sum_{s=1}^N \int d^3 x \left\lbrack 2\pi\, U_r^{\mu}
(\epsilon_{\mu\nu\gamma}\p^{\nu})\delta^{rs} U_s^{\gamma}  - \frac{g^2}{4 N} F_{\mu\nu} ( U_r ) M^{rs}
F^{\mu\nu} ( U_s ) \right\rbrack \label{sbmaisb}\ee

\be W_- = \sum_{r=1}^N \sum_{s=1}^N \int d^3 x \left\lbrack - 2\pi\, V_r^{\mu}
(\epsilon_{\mu\nu\gamma}\p^{\nu})\delta^{rs} V_s^{\gamma}  - \frac{g^2}{4 N} F_{\mu\nu} ( V_r ) M^{rs}
F^{\mu\nu} ( V_s ) \right\rbrack \label{sbmenosb}\ee

\no Where the matrix $M$, defined by $M^{rs}=1$ in all entries, can be made diagonal by an orthogonal
transformation: $U_r = \sum_{s=1}^N T_{rs} \tu_s \, ; \, V_r = \sum_{s=1}^N T_{rs} \tv_s$, with $(T^t T)_{rs} =
\delta_{rs}$. After such transformation we diagonalize the second term in (\ref{sbmaisb}) and (\ref{sbmenosb})
with $ T^t M T = diag(N,0,\cdots,0)$ without affecting the matrix structure of the first term which is already
diagonal. Explicitly,

\be W_+ = \int d^3 x \left\lbrack -\frac{g^2}{4} F_{\mu\nu}^2(\tu_1) + 2\pi
\epsilon_{\mu\nu\gamma}\tu_1^{\mu}\p^{\nu}\tu_1^{\gamma}  + 2\pi \sum_{r=2}^{N}
\epsilon_{\mu\nu\gamma}\tu_r^{\mu}\p^{\nu}\tu_r^{\gamma}\right\rbrack \label{sbmaisc} \ee

\be W_- = \int d^3 x \left\lbrack -\frac{g^2}{4} F_{\mu\nu}^2(\tv_1) - 2\pi
\epsilon_{\mu\nu\gamma}\tv_1^{\mu}\p^{\nu}\tv_1^{\gamma}  - 2\pi \sum_{r=2}^{N}
\epsilon_{\mu\nu\gamma}\tv_r^{\mu}\p^{\nu}\tv_r^{\gamma}\right\rbrack \label{sbmenosc} \ee

\no The only non vanishing eigenvalue $\lambda_1=N$ of the matrix $M$ corresponds to the normalized eigenvector
$\left(1,\cdots,1\right)/\sqrt{N}$. Therefore, $\tu_1^{\mu} = (U_1 + U_2 + \cdots U_N)^{\mu}/\sqrt{N}$ and
$\tv_1^{\mu} = (V_1 + V_2 + \cdots V_N)^{\mu}/\sqrt{N}$. The explicit form of the other fields $\tu_r
(U_1,\cdots , U_N),\tv_r (V_1,\cdots,V_N)$ with $r=2,\cdots,N$ depends explicitly on $N$. Since the last $N-1$
Chern-Simons terms in (\ref{sbmaisc}) and (\ref{sbmenosc}) have no particle content we just have  a couple of
massive modes of opposite helicities $\pm 1$ in the spectrum of $W_+$ plus $W_-$ in agreement with the soldered
theories (\ref{ls1}) or (\ref{ls2}). Apparently, the $N-1$ Chern-Simons terms have been canceled by the
soldering mechanism. Indeed, we can understand this point  and the appearance of the self-dual models
(\ref{ls1}) by applying the multiflavor soldering procedure directly in the new basis $\tilde{U}_r,\tilde{V}_r$.
Once again we choose $N=2$ for simplicity. In this case $\tu_1^{\mu} = (U_1^{\mu} + U_2^{\mu})/\sqrt{2} \, ; \,
\tv_1^{\mu} = (V_1^{\mu} + V_2^{\mu})/\sqrt{2} $ and $\tu_2^{\mu} = (U_1^{\mu} - U_2^{\mu})/\sqrt{2} \, ; \,
\tv_2^{\mu} = (V_1^{\mu} - V_2^{\mu})/\sqrt{2} $ Therefore, implementing the local transformations
(\ref{deltauv}) we have:

\bea \delta\tv_1 &=& \frac{(\alpha_1\eta_1 + \alpha_2\eta_2)}{\sqrt{2}} = \frac 12 (\alpha_1 + \alpha_2)\delta
\tu_1 + \frac 12 (\alpha_1 -
\alpha_2)\delta \tu_2 \label{deltac1} \\
\delta\tv_2 &=& \frac{(\alpha_1\eta_1 - \alpha_2\eta_2)}{\sqrt{2}} = \frac 12 (\alpha_1 - \alpha_2)\delta \tu_1
+ \frac 12 (\alpha_1 + \alpha_2)\delta \tu_2 \label{deltac2} \eea

\no In the case  $\alpha_1=\alpha_2=\alpha $ we have a ``parallel'' soldering while the case
$\alpha_2=-\alpha_1=-\alpha $ might be called a ``cross-soldering''. Next, we show that the parallel soldering
gives rise to (\ref{ls1}) while the cross-soldering originates (\ref{ls2}). In the first case we solder a
Maxwell-Chern-Simons (MCS) theory with another MCS one of opposite helicity which is known \cite{bk} to produce
a Maxwell-Proca theory, at the same time  the pure Chern-Simons (CS) theory of $W_+$ is soldered with the pure
CS theory of $W_-$ which cancel out completely as we show below. In the second case (cross-soldering) we solder
the MCS theory of $W_+$ with the pure CS theory of $W_-$ and vice-versa which produces (\ref{ls2}). In order to
clarify that point it is enough to consider the soldering of the following theories:

\bea W_{{\rm MCS}} &=& \int d^3 x \left\lbrack - \frac{g^2}4 F_{\mu\nu}^2 (U) + \frac a2
\epsilon_{\mu\nu\gamma}U^{\mu}\p^{\nu}U^{\gamma} \right\rbrack \label{wmcs}\\
W_{{\rm CS}} &=& -\frac b2\, \int d^3 x \, \epsilon_{\mu\nu\gamma}V^{\mu}\p^{\nu}V^{\gamma}\label{wcs} \eea

\no Where $a , b$ are constants. Imposing local invariance under $\delta U^{\mu} = \eta^{\mu} \, , \, \delta
V^{\mu} = \alpha\eta^{\mu}$ we have $\delta (W_{{\rm MCS}} + W_{{\rm CS}}) = \int d^3 x
J_{\mu\nu}\p^{\mu}\eta^{\nu} $ with $J_{\mu\nu} = - g^2 F_{\mu\nu}(U) + \epsilon_{\mu\nu\gamma}(a U^{\gamma} -
\alpha \, b \, V^{\gamma})$. For $\alpha = \pm \sqrt{a/b}$ we have $\delta J_{\mu\nu} = - g^2(\p_{\mu}\eta_{\nu}
- \p_{\nu}\eta_{\mu}) $. Therefore, by introducing an auxiliary anti-symmetric field such that $\delta
B_{\mu\nu} = - (\p_{\mu}\eta_{\nu} - \p_{\nu}\eta_{\mu})/2 $ we deduce:

\be \delta \left(W_{{\rm MCS}} + W_{{\rm CS}} + B_{\mu\nu}J^{\mu\nu}\right) = 2\, g^2\,  \int d^3 x\, B_{\mu\nu}
\, \delta B^{\mu\nu} \label{deltab} \ee

\no Consequently we have the soldered action:

\be W^{(S)} = W_{{\rm MCS}} + W_{{\rm CS}} + \int d^3 x \left\lbrack B_{\mu\nu}J^{\mu\nu} - g^2
B_{\mu\nu}B^{\mu\nu} \right\rbrack \label{wsb} \ee

\no Before we go further we notice that for $g=0$  the equations of motion for the auxiliary fields imply
$J_{\mu\nu}=0$ which leads to $V^{\mu} = \alpha U^{\mu}$, consequently recalling that $\alpha^2 = a/b$ we have
${\cal L}^{(S)} = \frac a2 \epsilon_{\mu\nu\gamma}U^{\mu}\p^{\nu}U^{\gamma}-\frac b2
\epsilon_{\mu\nu\gamma}V^{\mu}\p^{\nu}V^{\gamma} = 0$. Thus, the soldering of two CS theories, even with
different coefficients, cancel out completely and so the parallel soldering explains the final soldered theory
(\ref{ls1}). On the other hand, if $g\ne 0$, after the elimination of $B_{\mu\nu}$ through its equations of
motion, we see that the soldering of a MCS with a CS theory  produces a self-dual model written in terms of the
invariant combination $C_{\mu} = \alpha U_{\mu} - V_{\mu} $:

\be {\cal L}^{(S)} = - \frac b2 \epsilon_{\mu\nu\gamma}C^{\mu}\p^{\nu}C^{\gamma}  + \frac{a^2}{4 g^2}
C_{\mu}C^{\mu} \label{sd} \ee

\no By setting $a=b=4\pi$ we see that the cross-soldering correctly reproduces the couple of self-dual models
obtained in (\ref{ls2}). It is remarkable that for $g\ne 0$ the soldered theory (\ref{sd}) contains one massive
particle of mass $a^2/(2 \vert b \vert\, g^2)$ while before soldering we had a mass $ 2 a/g^2$. Such change of
mass by means of soldering appeared before in the fusion of two chiral Schwinger models (CSM) of opposite
chiralities, which have altogether two massless particles in the spectrum, into one vector \cite{abw} Schwinger
model (VSM) or one axial \cite{dda} Schwinger model (ASM) which contain one massive particle in the spectrum.
The similarities with the two-dimensional case will be further explored in the next section.

\subsection{Dual theories via generalized soldering}

In \cite{dda} we have introduced the parameter $\alpha$ in the soldering procedure (generalized soldering) and
shown that we can solder two CSM of opposite chiralities into one VSM ($\alpha =1$) or one ASM ($\alpha=-1$):

\begin{eqnarray}
\mathcal{L}_{{\rm VSM}} &=&{\frac{1}{2}}\,(\partial _{\mu }\,\Phi )^{2}\,+\,e\,\varepsilon ^{\mu \nu }\,\partial
_{\mu }\,\Phi
\,A_{\nu }\,,  \label{vsm} \\
\mathcal{L}_{{\rm ASM}} &=&\frac{1}{2}\left(
\partial ^{\mu }\Phi +e\,A^{\mu }\right) ^{2}.  \label{csm}
\end{eqnarray}

\no Although the models (\ref{vsm}) and (\ref{csm}) look quite different, after adding a Maxwell term, already
present before soldering in the chiral Schwinger models, and  integrating in the path integral over the
soldering field $\Phi$  we have the same final effective theory with one massive particle in the spectrum, i.e.,
$\mathcal{L}_{eff}\left[ A_{\mu }\right] =- (1/4)F^{\mu \nu } \left( \Box +e^{2}/\pi \right)\Box ^{-1} F_{\mu
\nu }$. This suggests that the ASM and the VSM are  dual versions of the same theory. In fact, see e.g.
\cite{Anacleto2}, a one-step Noether gauge embedding procedure has been developed and applied on a variety of
cases in order to obtain dual versions of a given theory. In particular, starting from the self-dual model
\cite{tpn} one ends up with the MCS model thus reproducing a well known duality relation. Now we show that
applying a similar procedure one can go from the VSM to the ASM. Namely, starting from $\mathcal{L}_{{\rm VSM}}
$  and imposing symmetry under the local transformations $\delta\Phi = e \, \eta \, , \, \delta A_{\mu} = -
\p_{\mu} \eta $ we have $\delta \mathcal{L}_{{\rm VSM}} = J_{\mu} \p^{\mu} \eta $ where $ J_{\mu} = e
\left\lbrack \p_{\mu}\Phi - \epsilon^{\nu}_{\mu}\p_{\nu} \Phi + e \epsilon_{\mu\nu}A^{\nu}  \right\rbrack $.
Thus, introducing an auxiliary field such that $\delta B_{\mu} = - \p_{\mu} \eta $ we have $\delta (
\mathcal{L}_{{\rm VSM}} + B_{\mu} J^{\mu} ) = e^2 B_{\mu}\p^{\mu}\eta = \delta \left(- e^2 B_{\mu}B^{\mu}/2
\right) $. Consequently, defining ${\cal L} = \mathcal{L}_{{\rm VSM}} + B_{\mu} J^{\mu}  + e^2 B_{\mu}B^{\mu}/2
$ we have $\delta {\cal L}=0$. Eliminating the auxiliary field using its equation of motion we have finally the
ASM :

\be {\cal L} = \mathcal{L}_{{\rm VSM}} - \frac{J_{\mu}J^{\mu}}{2
e^2} = \frac 12\left(
\partial ^{\mu }\Phi +e\,A^{\mu }\right) ^{2} \ee

\no Apparently, the generalized soldering $\alpha=\pm 1$ gives rise to dual theories but if we try to use it to
fuse two MCS theories of opposite helicities we only have a Maxwell-Proca theory with no possibilities of having
a dual formulation. It is remarkable that if we add in both MCS theories an independent Chern-Simons term, which
works like a spectator in the original model and does not change anything in the physics of the model, we have
the possibility of performing a cross-soldering between the MCS of one theory and the CS term of the other
theory and vice-versa thus leading to a couple of self-dual models which is dual to a Maxwell-Proca theory. In
conclusion, with the help of the spectators Chern-Simons terms we can still produce dual theories in $3D$ by
means of the generalized soldering procedure. Thus, completing the analogy with the $2D$ case where no spectator
field is necessary.

\section{Conclusion}

With or without a coupling to a gauge field it is possible to
achieve a partial bosonization of the Thirring model in $D=2+1$ by
simply integrating perturbatively over the two components
fermions, see \cite{fs,kondo,bm,ddh}. Although the detailed
calculations are different, they all start with the computation of
the vacuum polarization diagram that gives in general a nonlocal
effective bosonic action which becomes local at the limit
$m\to\infty $. Some approaches use a small coupling expansion
\cite{kondo,bm} while others make use of a $1/N$ expansion
\cite{fs,ddh} which one may find more appropriate due to the power
counting nonrenormalizability of the Thirring coupling in $D=2+1$.
The authors of \cite{fs} have arrived at a self-dual or an
anti-self-dual model depending on the helicity of the fermions. In
order to keep a one to one correspondence in the number of degrees
of freedom we have found natural \cite{ddh} to introduce $N$
bosonic vector fields, in correspondence with each $U(1)$ current
$\bar{\psi}_{r} \gamma_{\mu} \psi_{r}$ , $ r=1,\cdots,N $.
Consequently, from the two point correlators of those $U(1)$
currents we obtain \cite{ddh} the coupled Maxwell-Chern-Simons
actions given in (\ref{sbmais}) and (\ref{sbmenos}) respectively
for fermions of $+1/2$ and $-1/2$ helicities. A simple field
redefinition, see section 3, shows that (\ref{sbmais}) and
(\ref{sbmenos}) are physically equivalent to two MCS theories of
opposite helicities which are known \cite{dj} to be dual to the
self-dual and anti-self-dual models, therefore, although the
number of bosonic flavors is different, the truncation of the
fermionic determinant to the quadratic order in both approaches
\cite{fs,ddh} has led us to a dual Lagrangian with the same
particle content. Such conclusion  has been suggested to us by the
twofold soldering results of the second section which has
explicitly provided the missing interference term which takes us
from one formulation to the other one when we combine both
helicities altogether. We have thus a closer connection between
two different perturbative approaches used in the literature to
partially bosonize the $3D$ Thirring model.

It is remarkable that we have derived the interference term by
introducing a spectator field (non-interacting Chern-Simons field)
which amounts to a normalization factor in the fermionic
determinant and does not add anything in the physics of the
original MCS theories, but plays an active role in the
cross-soldering procedure.

A comparison between the $D=2$ and $D=3$ applications suggests
that the generalized soldering could be used to provide dual
theories.

At last, although we have been dealing throughout  this work with
quadratic theories, the use of a Noether gauge embedding procedure
in \cite{Anacleto2,jpa} has furnished interesting results for
truly interacting theories. Since such embedding is in the heart
of the soldering procedure we hope to generalize our results to
interacting non-trivial cases. It would be interesting also to
generalize the soldering of the non-abelian 2D theories of
\cite{abw} to the 3D case.

\section{Appendix}

In this appendix we obtain the interference Lagrangian for multiflavor soldering in the case of $N=3$ flavors.
In such case the piece of ${\cal L}^{(S)}$ which depends on the auxiliary fields is given by:

\bea {\cal L}^{(S)}_{N=3} (B^{\mu\nu}_r) &=& B_1 \cdot J_1 +  B_2 \cdot J_2 + B_3 \cdot J_3 - \frac{2 g^2}3
\left\lbrack B_1\cdot B_1 +   B_2\cdot B_2 +  B_3\cdot B_3 \right. \nn \\
 &+& \left. (1 + \alpha_1\alpha_2) B_1 \cdot B_2  + (1 +
\alpha_1\alpha_3) B_1 \cdot B_3 + (1 + \alpha_2\alpha_3) B_2 \cdot B_3  \right\rbrack \label{l3a} \eea

\no Since two of the three $\alpha_i$ above must be equal we choose without loss of generality
$\alpha_1=\alpha_2=\alpha $. In this case we have

\bea {\cal L}^{(S)}_{N=3} (B^{\mu\nu}_r) &=& B_- \cdot J_- +  B_+ \cdot J_+ + B_3 \cdot J_3 \nn \\
&-& \frac{2 g^2}3 \left\lbrack B_+\cdot B_+ + B_3\ \cdot B_3 + (1 + \alpha\alpha_3) B_+ \cdot B_3
\right\rbrack \label{l3b} \eea

\no Where $B_{\pm}^{\mu\nu} = B^{\mu\nu}_1 \pm B^{\mu\nu}_2 \, , \, J_{\pm}^{\mu\nu} = \left( J^{\mu\nu}_1 \pm
J^{\mu\nu}_2 \right)/2 $. Now we have two possibilities: either $\alpha_3 = \alpha $ or $\alpha_3 = - \alpha $.
In the first case we get:

\bea {\cal L}^{(S)}_{N=3} (B^{\mu\nu}_r) &=& B_- \cdot J_- + \frac{(B_+ -
B_3)\cdot (J_+ - J_3)}2 \nn \\
&+& \frac{(B_+ + B_3)\cdot (J_+ + J_3)}2 - \frac{2 g^2}3 (B_+ + B_3)^2 \eea

\no Therefore, the elimination of $B^{\mu\nu}_{-}$  and the combination $B^{\mu\nu}_{+} - B^{\mu\nu}_{3}$ will
lead us to the identification $J_{\mu\nu (1)}=J_{\mu\nu (2)}=J_{\mu\nu (3)}\equiv J_{\mu\nu (\alpha)}$.
Consequently, $(J_{\mu\nu (+)} + J_{\mu\nu (3)})/2=J_{\mu\nu (\alpha)}$. Thus, using the equation of motion for
the combination $B^{\mu\nu}_{+} + B^{\mu\nu}_{3}$ we end up with the interference Lagrangian density ${\cal
L}_I^{(2)} = (3/8 g^2) J_{\alpha}\cdot J_{\alpha}$.

On the other hand, in the second case where $\alpha_3=-\alpha$ in (\ref{l3b}) we have :

\bea {\cal L}^{(S)}_{N=3} (B^{\mu\nu}_r) &=& B_- \cdot J_- + B_+ \cdot J_+  + B_3 \cdot J_3 - \frac{2 g^2}3 (
B_+ \cdot B_+ + B_3\cdot B_3) \eea

\no Now, the elimination of $B^{\mu\nu}_{-}$ implies $J_{\mu\nu (1)}=J_{\mu\nu (2)}= J_{\mu\nu (\alpha)}$ and
$J_{\mu\nu (3)}= J_{\mu\nu (-\alpha)}$. After the use of the equations of motion of $B^{\mu\nu}_{+}$ and $B^
{\mu\nu}_3$ one obtains a different interference Lagrangian density ${\cal L}_I^{(1)}= (3/8 g^2)
(J_{\alpha}\cdot J_{\alpha} + J_{-\alpha}\cdot J_{-\alpha})$. Both ${\cal L}_I^{(1)}$ and ${\cal L}_I^{(2)}$
keep their form for arbitrary $N$.

\section{Acknowledgments}

This work is partially supported by CNPq (Brazilian agency).

\end{document}